\documentclass[a4paper]{aa} 
\usepackage[dvips]{graphics}
\usepackage{graphicx}
\usepackage{xspace}
\usepackage{amssymb,amsmath} 
\newcommand{\lo}     {\stackrel{<}{_\sim}\,}

\newcommand{\ascc}{\mbox{ASCC-2.5}\xspace}
\newcommand{\clucat} {\mbox{COCD}\xspace}

\newcommand{\oline}[1]{\overline{#1}}
\renewcommand{\ni}{\noindent}
\usepackage{txfonts}
\begin{document}

\titlerunning{Astrophysical parameters of Galactic open clusters}
\authorrunning{Kharchenko et al.}

\title{Astrophysical Parameters of Galactic Open Clusters}

\author{N.V.~Kharchenko \inst{1,2,3} \and
	     A.E.~Piskunov	\inst{1,2,4} \and
        S.~R\"{o}ser    \inst{1},
     	  E.~Schilbach	\inst{1} \and
     	  R.-D.~Scholz	\inst{2}
       }

\offprints{R.-D.~Scholz}

\institute{Astronomisches Rechen-Institut, M\"onchhofstra{\ss}e 12-14,
             D--69120 Heidelberg, Germany\\
             email: nkhar@ari.uni-heidelberg.de, piskunov@ari.uni-heidelberg.de, roeser@ari.uni-heidelberg.de, elena@ari.uni-heidelberg.de
	 \and
     	  Astrophysikalisches Institut Potsdam, An der Sternwarte 16,\\
     	     D--14482 Potsdam, Germany\\
             email: nkharchenko@aip.de, apiskunov@aip.de, rdscholz@aip.de
         \and
          Main Astronomical Observatory, 27 Academica Zabolotnogo Str.,
             03680  Kiev, Ukraine\\
             email: nkhar@mao.kiev.ua
           \and
           Institute of Astronomy of the Russian Acad. Sci.,
           48 Pyatnitskaya Str., Moscow 109017, Russia\\
           email: piskunov@inasan.rssi.ru
	  }

\date{Received ... December 2004; accepted ...}

\abstract{ \ni  We present a catalogue of astrophysical data of 520 Galactic
open clusters. These are the clusters, for which at least three most probable
members (18 on average) could be identified in the \ascc, a  catalogue of stars
based on the Tycho-2 observations from the Hipparcos mission. We applied
homogeneous methods and algorithms to determine angular sizes of cluster cores
and coronae, heliocentric distances, mean proper motions, mean radial
velocities and ages. For the first time we derive distances for 200 clusters,
radial velocities for 94 clusters and determine ages of 196 clusters. This
homogeneous new parameter set is compared with earlier determinations. In
particular, we find that the angular sizes were systematically underestimated
in the literature.
}


\maketitle

\section{Introduction}

Although open clusters are typical representatives of the Galactic disk
population, a systematic investigation of their nature, size, number of members
and age is hampered by the inhomogeneity of the data. On one hand, general
bibliographic catalogues derived from reviews of the literature are available.
Examples of this category are the catalogues by Lyng{\aa}~(\cite{lyn87}),
Ruprecht et al.(~\cite{rupr}), Dias et al.~(\cite{dlam}), and the data base
WEBDA\footnote{\texttt{http://obswww.unige.ch/webda/}} of Mermilliod. In these
collections general parameters of the clusters are taken from the literature as
published. These collections are indispensable sources for work in this field.
However, their parameters cannot be used for systematic studies or
intercomparisons between clusters. On the other hand,  there are uniform lists
of a few hundreds of clusters,  the parameters of which were derived either
from coherent photometric studies (e.g. Becker and Fenkart~\cite{baf71}, Janes
and Adler~\cite{janad82}, Loktin et al.~\cite{lok01},
Tadross~\cite{tad}), or from their kinematics (e.g. Baumgardt et
al.~\cite{baum00}, Dias et al.~\cite{dla1}, \cite{dla2}).

A few years ago we started the project to determine membership and principal
parameters of open clusters by use of objective methods and algorithms. In
order to get such an unbiased view onto the system of open clusters of our
Galaxy, one has to start with an homogeneous sky survey obtained with no
particular emphasis on open clusters. One result of ESA's Hipparcos mission,
the Tycho-2 catalogue  (H{\o}g et al.~\cite{tyc2}), fulfills this necessary
criterion,  but its contents alone were not sufficient for our purpose.

So, as a first step, a catalogue of 2.5 million stars with proper motions in
the Hipparcos system and $B$, $V$ magnitudes in the Johnson photometric system
was compiled and supplemented with spectral types and radial velocities if
available. The resulting catalogue, the All-Sky Compiled Catalogue
of 2.5 Million Stars (\ascc, Kharchenko~\cite{kha01}) can be
retrieved from the CDS
\footnote{\texttt{ftp://cdsarc.u-strasbg.fr/pub/cats/I/280A}}, a detailed
description of the catalogue can be found in that paper or in the corresponding
ReadMe file at the CDS.

In a second step, we used the \ascc to identify known open clusters and compact
associations listed in the Lund Catalogue of Open Clusters
(Lyng{\aa}~\cite{lyn87}), the Optically Visible Open Clusters Catalog (Dias et
al.~\cite{dlam}, referred  hereafter as DLAM), and the Catalogue of Star
Clusters and Associations (Ruprecht et al.~\cite{rupr}). Applying an iterative
procedure of cluster membership determination based on proper motion,
photometric and spatial criteria to the \ascc data, we could identify 520 of
about 1700 known clusters. After re-estimating the  positions of cluster
centres and cluster sizes, about 150\,000 \ascc stars  were selected in
these 520 cluster areas, and  membership probabilities were computed for each
star in this list which we called the Catalogue of Stars in Open Cluster Areas
(CSOCA). The procedure of cluster membership determination is briefly described
below, a detailed information is given in Kharchenko et al.~(\cite{starcat}),
referred hereafter as Paper~I. CSOCA is a catalogue of stars which now gives us 
the basis for deriving uniform structural (location, size), kinematic (proper
motions and spatial velocities) and evolutionary (age) parameters for open
clusters in the wider neighbourhood of the Sun. The results are published in
the Catalogue of Open Cluster Data (COCD) and the Open Cluster Diagrams Atlas
(OCDA).

The present paper is the third step in our long-term project, and it is
structured as follows: in Sec.~\ref{memlist_sec} we briefly describe the basic
data and the procedure of member selection we developed. In
Sec.~\ref{struct_sec} the parameters describing the spatial-structure of the
clusters are discussed. Sec.~\ref{kin_sec} is devoted to the kinematics of the
clusters. In Sec.~\ref{age_sec} we describe the method applied for deriving
ages of the clusters. Concluding remarks are given in Sec.~\ref{concl_sec}. In
the appendices we describe the format of COCD and OCDA.

\section{The cluster list, and cluster membership}\label{memlist_sec}

The input data for this study come  from the \ascc and cover 513
open clusters and 7 compact associations (i.e. the associations with a
sufficiently high projected density of stars, so that  the developed general
procedure of member selection works properly). Due to a relatively bright
limiting magnitude ($V\approx 14$ mag) of the \ascc, our sample does
not include faint (and generally remote or highly obscured) open clusters.
Also, the two nearest extended clusters, the Hyades and Cr~285 (the UMa
cluster), are missing in our list since they require a  special and more
sophisticated technique of membership determination. However, from comparison
with DLAM, our sample is sufficiently complete for clusters up to 1 kpc and
can be used for  the study of cluster parameters and properties of the local
population of  open clusters. 

For each cluster, the membership determination is based on a comprehensive
common  analysis of several diagrams derived with \ascc data: a sky chart
referring to the cluster, the radial distribution $F(r)$ of the projected stellar
density, a vector point diagram (VPD) of the proper motions, the magnitude dependence
of the proper motion components, and a colour-magnitude diagram (CMD). 

\begin{figure}[t] \resizebox{\hsize}{!}
{\includegraphics[bbllx=35,bblly=45,bburx=565,bbury=795,angle=270,clip=]
{figs/stock2.ps.gz}} 
\caption{Illustration of the member-selection algorithm for the open cluster
Stock~2. Arrows show the sequence of the basic selection phases presented by
rectangles. Panel~(a) is a sky map of the cluster neighbourhood.  Panel~(b)
shows radial profiles of the projected density, panel~(c) is the
colour-magnitude diagram, panel~(d) is vector point diagram,  and panels~(e)
and (f) are  "magnitude equations" ($\mu_{x,y}-V$ relations;  they are useful
as a proper motion check since proper motions of cluster  members should not
depend on their magnitudes). Additional explanations for each panel are given
in Appendix~\ref{atl_sec}.}\label{memb_fig}
\end{figure}

As a first approximation of the member selection algorithm applied, we consider
all \ascc stars within an area of 1$\times$1 sq. deg around the cluster centre
with coordinates taken from DLAM. Using proper motion and photometric criteria,
we then separate field stars and cluster members and adopt the point of maximum
surface density of cluster  members as the new approximation of the location of
the cluster centre. From the analysis of the surface cluster density $F(r)$, a
projected radius of the cluster is then derived. The resulting spatial
parameters are used for the next steps in the iteration. The iterations are
stopped when the cluster member list does no longer change. As a rule, two
iterations are sufficient to establish cluster membership and determine the
structural and kinematic parameters. However, some clusters with sparse
structure or initially erroneous central coordinates require three and more
iterations. The member selection pipeline is illustrated in Fig.~\ref{memb_fig}
for the case of the open cluster Stock~2 (see Paper~I for more details).

For stars, the probability of belonging to a cluster is calculated as a measure
of a deviation either from the cluster mean proper motion (kinematical
probability), or from the Main Sequence (MS) edges (photometric probability).
Stars deviating from the reference values by less than one $\sigma$ $rms$ are
classified as most probable cluster members (1$\sigma$-members, i.e., with a 
membership probability $P \ge 61$\%). Those falling in semi intervals
[1$\sigma$,2$\sigma$) or [2$\sigma$,3$\sigma$) are considered as possible
members ($P=14-61$\%) or  possible field stars ($P=1-14$\%), respectively.
Stars with deviations larger than 3$\sigma$ are regarded as definite field
stars ($P < 1$\%).  As a rule, all cluster parameters were determined by use of
data on the most  probable cluster members.

\section{Spatial structure parameters of the clusters}~\label{struct_sec}

\subsection{Distances and extinction }\label{dist_sec}

Although about  9\,700 of the CSOCA stars with cluster membership probabilities
higher than 14\,\% (i.e. 1$\sigma$- or 2$\sigma$-members) have trigonometric 
parallaxes measured by Hipparcos/Tycho, these parallaxes are significant  
(i.e. $\pi > 3\sigma_{\pi}$) for only 640 of these stars.  Reliable
Hipparcos-based distances to a few nearby clusters have already been 
determined by Robichon et al.~(\cite{rob99}). But for the vast majority of the 
clusters under study, trigonometric parallaxes are not sufficiently accurate or
not available at all. Therefore, we are forced to use indirect estimates of
distances.

The photometric approach of simultaneous determination of  distance and
interstellar extinction requires at least 3-colour photometry which is lacking
in the \ascc. Therefore, for 255 clusters  we took the data from a  list
derived and newly revised by Loktin et al.~(\cite{lok01}, \cite{lok04}) (LGM
hereafter) who estimated distances and interstellar extinctions from
homogeneous photometric parameters.  Further, we include data of Robichon et
al.~(\cite{rob99}) -- 8 clusters, Lyng{\aa}~(\cite{lyn87}) -- 19 clusters, 
DLAM -- 31 clusters, de Zeeuw et al.~(\cite{dez})  -- 1 association, Ruprecht
at al.   (\cite{rupr}) -- 6 associations.

For the remaining 200 clusters we determined or revised\footnote{For several
clusters the published  distance estimates do not fit the observed CMDs and
were, therefore,  recalculated (e.g. for the Pleiades we obtained a distance of
130~pc instead of 118~pc in Robichon et al.~(\cite{rob99}) and 150~pc in LGM).}
cluster distances and reddening by use of supplementary data on spectral
classes of the most probable members available from the \ascc and the Tycho-2
Spectral Type Catalog (Wright et al.~\cite{ty2sp}). Further, we adopted
Schmidt-Kaler's~(\cite{schkal}) Zero Age Main Sequence (ZAMS) for the Main
Sequence (MS) fitting, and  $A_V =3.1 \times E(B-V)$ for computing the total
interstellar extinction and colour-excess. The spectral class-colour/absolute
magnitude calibration was based on  Strai\v{z}ys~(\cite{strai}). 

The references for adopted extinction and distances are provided in the \clucat
for each cluster.

\begin{figure}[t]
\resizebox{\hsize}{!}
{\includegraphics[bbllx=221,bblly=100,bburx=482,bbury=660,angle=270,clip=]
                 {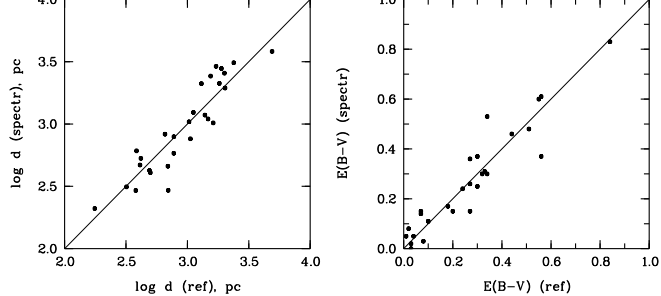}}
\caption{Comparison of distances $d$ and colour-excesses \mbox{$E(B-V)$} derived
by Loktin et al.~(\cite{lok01}, \cite{lok04}) from 3-colour photometry
with values determined on the basis of spectral classification. }
\label{cdist_fig}
\end{figure}

In order to check the reliability of the determined distances and extinction,
we applied our procedure to a few clusters selected arbitrary from LGM  and
distributed over a wide range of distances. The results are compared with the
original LGM data and  given in  Fig.~\ref{cdist_fig}. We may conclude that
distances and colour excesses derived by the spectral  method coincide well
with the published LGM data based on 3-colour photometry.

\subsection{Angular  sizes}

Spatial parameters of the clusters were determined from profiles of stellar
density derived from star counts. For each cluster, the counts were carried out
in concentric  circles around the cluster centre. The centres were determined
as the points of maximum surface density of the cluster members (see
Sec.\ref{memlist_sec}).

As a rule, the differences between the determined coordinates of cluster 
centres and those listed in DLAM are small. Only for 21 clusters they exceed
$0\,\fdg3$ either due to errors in the cluster coordinates in old catalogs
(particularly, in the Collinder list), or due to problematic definition of
centres of large open clusters like Mel 20 ($\alpha$ Per), Mel 111 (Coma
Berenicis), poor stellar groups (some objects from Platais et
al.~\cite{pla98}), and cluster-like associations as, for example, Sco~OB4. 

A general model describing the structural parameters of clusters was developed
by  King~(\cite{king62}) and successfully applied to globular clusters. Since
a  typical open cluster has a relatively small number of members and the
spatial distribution of members is not regular, it is difficult to find a
formal model which would, on one hand, describe structural details of a given
cluster and, on the  other hand, would be valid in general. As a compromise, we
assumed a centrally  symmetric distribution of the cluster members and
considered only two structural components, a core with a radius $r_1$ and a
corona with a  radius $r_2$. The core radius corresponds to a distance where
the decrease of stellar surface density stops abruptly. The corona radius (i.e.
the actual radius of a cluster) is defined as the distance from the cluster
centre where the surface density of stars becomes equal to the average density 
of the surrounding field (see Fig.~\ref{memb_fig}(b)). For each cluster, core
and corona radii were checked by visual inspection, and the constrains set by
the corresponding VPD and CMD were always taken into account. Even though this
approach is somehow subjective and time consuming, it can be  applied to poorer
clusters and thereby considerably expand the sample studied.

\begin{figure}[t] 
\resizebox{\hsize}{!}
{\includegraphics[bbllx=75,bblly=205,bburx=500,bbury=520,clip=]
                 {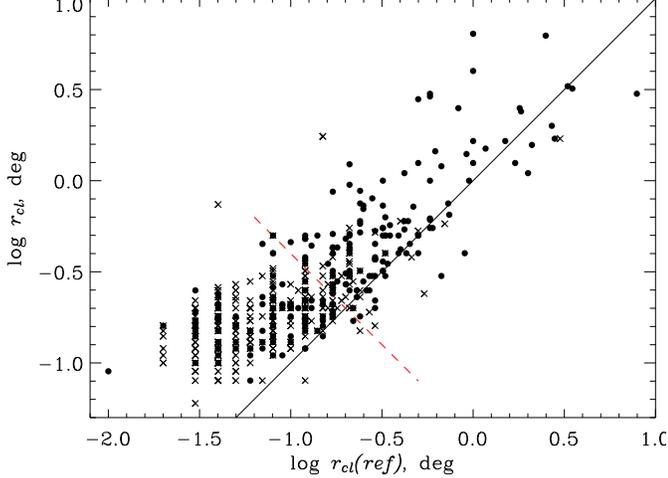}}
\caption{Comparison of derived angular radii with data compiled by Dias et
al.(~\cite{dlam}). The solid line is the locus of equal radii. Crosses mark
clusters with distances $d \ge 1$~kpc, whereas the dots are nearer clusters.
The dashed line separates ``large'' and ``small'' clusters.
}\label{racom_fig} 
\end{figure}

For each cluster, the surface density was computed in concentric strips  of
$0\,\fdg05$  up to 10~degrees from the cluster centre and for three different
stellar samples: (i) all stars, (ii) cluster members with probabilities $P \ge
1$\%, and (iii) cluster members with probabilities $P \ge 61$\%. Nevertheless,
the distribution of $1\sigma$ members ($P \ge 61$\%) was the decisive factor
for  the determination of the cluster radius. On average, a core radius is
about 2.5 times smaller than a corona radius. Fig.~\ref{memb_fig} (b)
illustrates the degree of variation of F(r) with radial distance in the case of
open cluster Stock~2.

For 515 clusters in common, Fig.~\ref{racom_fig} compares the angular radii of 
cluster  coronas determined in this paper and the corresponding  data compiled 
by DLAM\footnote{The DLAM list includes a number of clusters with cluster sizes
taken directly from the Lyng{\aa} catalogue}. Independent of the distances, the
published radii are in average lower (by about a factor of 1.5 for ``large''
clusters and by a factor of 2.5 for ``small'' clusters).  There can be several
reasons for this bias. Often, the cluster sizes are empirically derived from
star counts or even from visual inspection of cluster areas,  without previous
membership determination. The area of concentration of the brightest stars
usually defines the cluster size, and fainter cluster members are lost in the
rich fore- or background. In this case, one should speak rather on the ``core
radius'' than on ``cluster radius''. Sometimes, a cluster size is estimated
from dedicated observations in small sky fields.  The investigations are
limited by the detector used in the study (e.g., photographic plate or CCD
frame), and therefore, give rather underestimated cluster radii. Finally,
adopting the cluster sizes from the Lund catalogue, one should not forget that
Lyng{\aa}(~\cite{lyn87}) himself aimed to  publish the sizes of the cluster
cores. 

Our determinations of cluster sizes are not restricted by limited sky areas,
and  the information on membership is used. Nevertheless, the sizes derived are
based on counts of relatively bright stars 
($V \lesssim 12$), and thus they can be influenced by the mass 
segregation effect (see e.g. Raboud \& Mermilliod~\cite{rab98},
or Kharchenko et al.~\cite{khea03}). Therefore our data should be considered as
lower limits of the actual cluster sizes.

\section{Cluster kinematic parameters}\label{kin_sec}

\begin{figure}[t] 
\resizebox{\hsize}{!} 
{\includegraphics[bbllx=100,bblly=50,bburx=550,bbury=655,angle=270,clip=]
                 {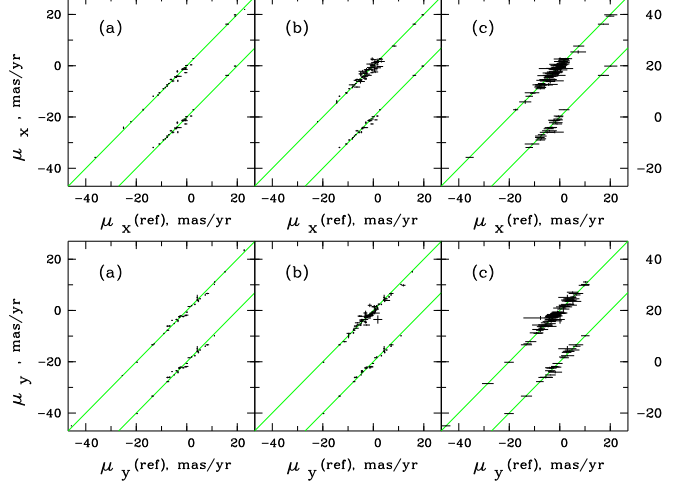}}
\caption{Comparison of cluster proper motions derived in this study from \ascc
data with results of (a) Robichon et al.~(\cite{rob99}), (b) Baumgardt et
al.~(\cite{baum00}), and (c) Dias et al.~(\cite{dla1}, \cite{dla2}). Bars show
$rms$-errors. The straight lines are the loci of equal proper motion 
components. The upper lines and left axes refer to clusters included in our
sample and in the samples of comparison. The lower lines and right axes are for
35 clusters common to all samples.
}\label{pmcom_fig}
\end{figure}

\subsection{Proper motions}

For each cluster, the mean components $\oline{\mu}_{x,y}$ of common proper
motion were computed from proper motions of the most probable members ($P \ge
61$\%). The number of such stars varies from 3 to 178, with an average of 18
stars per cluster. For 91\% of the clusters, the proper-motion components
$\oline{\mu}_{x,y}$ are determined with standard errors of less than $\pm
1$~mas/yr (for 42\% - less  than $\pm 0.5$ mas/yr). The proper motions are
derived directly in the Hipparcos system, for 301 clusters -- for the 
first time (see also Kharchenko e.a.~\cite{khea03}).

Fig.~\ref{pmcom_fig} compares the cluster proper motions derived in this study
with the corresponding findings of  Robichon et al.(\cite{rob99}), Baumgardt et
al.(\cite{baum00}), Dias et al.~(\cite{dla1}, \cite{dla2}) which are based on
the data from the Hipparcos and Tycho-2 catalogues. Although different
techniques of member selection were applied, the mean proper motions of the
common clusters agree quite well. Even clusters with small proper motions
coincide within a few mas/yr. This is remarkable because in these cases 
cluster members are difficult to separate from field stars.

\subsection{Radial velocities}

It is well known that our knowledge on  radial velocities (RV)  of open
clusters is much poorer than on proper motions. According to DLAM, RVs have
been published for only 240 of about 1700 known  clusters. Moreover, these 
data are very inhomogeneous: sometimes the RV of a cluster is taken from
measurements of only one star, sometimes the $rms$ errors reach 30~km/s, and
some authors did not give any information on the accuracy at all. 
Fortunately, the situation will be considerably improved when the 
RAVE programme (Steinmetz~\cite{rave03}) will be completed in the next years,
including some dedicated observations in the Galactic plane.

Radial velocities for only 196 clusters of our sample are listed in DLAM, the
Lund Catalogue and Ruprecht et al.~(\cite{rupr}). In order to update the RVs of
these clusters, we cross-identified (Kharchenko  et al.~\cite{kharv04}) the
\ascc with the General Catalogue of Radial Velocities (Barbier-Brossat \&
Figon~\cite{gcrv}). On the basis of our membership  determination, we were able
to revise the RVs for 159 clusters. Additionally, for 94 clusters, RVs have
been determined for the first time.  So, we can now publish RVs of 290 open
clusters  of our sample.

\begin{figure}[t] \resizebox{0.95\hsize}{!} 
{\includegraphics[bbllx=80,bblly=80,bburx=545,bbury=563,angle=270,clip=]
                 {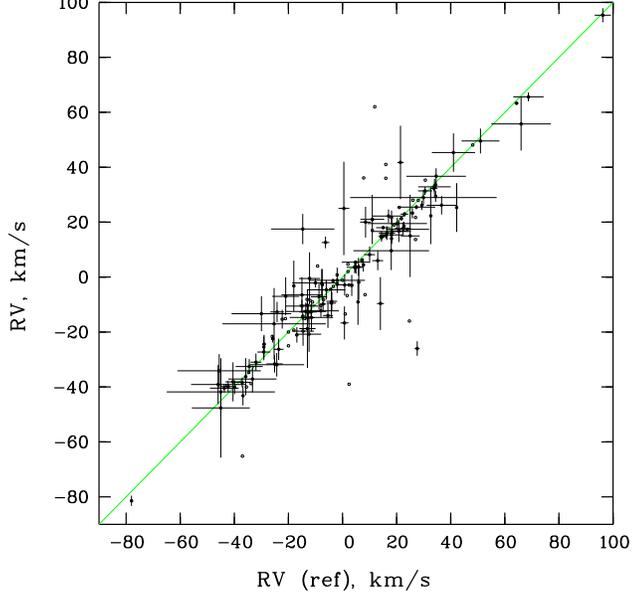}}
\caption{Comparison of cluster radial velocities derived in
this study with data published in Lyng{\aa}~(\cite{lyn87}),
Ruprecht et al. (\cite{rupr}), and DLAM. If the mean RV of a cluster is
determined from more than two members, error bars are shown.}\label{vrcom_fig}
\end{figure}

In Fig.~\ref{vrcom_fig}  the revised RVs for 160 clusters are compared with the
published data collected in DLAM. The mean difference between the ``old'' and
``new'' RVs of these clusters is $\overline{RV_ {ref} - RV}= 0.36 \pm 0.88$
km/s.

\section{Cluster ages}\label{age_sec}

In the preliminary version of the \clucat (Kharchenko et al.~\cite{khea03}), 
cluster ages were taken from the literature. Since different authors used
different methods of age estimates, these ages represented an inhomogeneous 
set of data in contrast with the homogeneity of other cluster parameters in 
the catalogue of Kharchenko et al.~(\cite{khea03}).  In the current version of
the \clucat we implement our own isochrone-based procedure of age
determination, which provides  a uniform scale of ages for all clusters. 

The contents of the \clucat catalogue put several constraints on the
theoretical models which can be used for the age estimates. Since a wide span
of cluster ages is expected, the CMDs of many clusters in our sample should
present the evolved portions of the upper Main Sequence. On the other hand, the
Pre-MS branches, observed at relatively  faint absolute magnitudes should be
seen in the \clucat cluster diagrams in rare  cases of young and nearby
clusters only.  Taking the above into account, we have been  concentrating on
the  Post-MS isochrones. Due to the rather bright limiting magnitude of the
\ascc,  our  sample is biased towards local clusters with typical distances
less than 1.5~kpc from the Sun. This means that for our sample, no considerable
metallicity trends   due to the radial gradient of [Fe/H] in the Galactic disk
is expected, and we can  limit ourselves to the consideration of solar
metallicity isochrones only.

\begin{figure*}[t]
\resizebox{!}{7cm}
{\includegraphics[bbllx=55,bblly=240,bburx=530,bbury=610]
                  {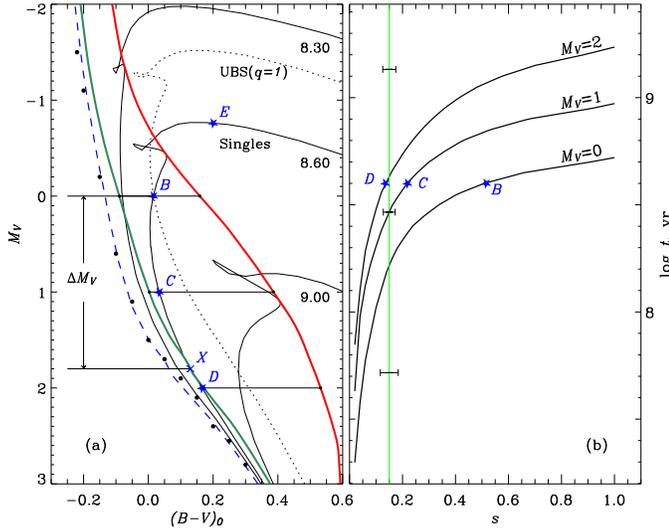}}
\parbox[b]{9cm}{
\caption{Illustration of the adopted algorithm for determination of cluster
ages. In Panel~(a), the thin curves are isochrones for three different ages
($\log t$ = 8.30; 8.60; 9.00). The dotted curve is the locus of unresolved
binaries with mass ratio $q=1$ for the age $\log t = 8.60$. 
The thick dashed curve
is the adopted ZAMS where the dots mark the empirical ZAMS of Schmidt-Kaler
(\cite{schkal}). The thick curves to the right of the ZAMS are the blue
($s=0.15$) and red (i.e., TAMS, $s=1.00$) borders of the evolved MS. The
horizontal lines indicate the EMS spread at $M_V$ = 0; 1; 2. The age variations
along these lines are shown in Panel~(b) as  functions of the  evolutionary
status $s$ (the blue edge of the evolved MS is marked by the vertical line at
$s=0.15$). Photometric errors in $(B-V)$ of 0.01~mag are converted to
corresponding errors in $s$ and shown  as bars. Stars marked by letters
$B,C,D,E$ are hypothetical cluster members, where $B$ is the brightest MS star.
$X$ marks the isochrone base point defined as the cross-section of the
corresponding isochrone with the  blue edge of the evolved MS.  $\Delta M_V$ is
the magnitude range  adopted for the cluster age evaluation. Whereas stars $B$
and $C$ will be included by the algorithm in the age determination, stars $D$
and $E$  will be rejected. 
}\label{agev_fig}
}
\end{figure*}

\subsection{Input data}

There are three recently published data sets of Post-MS isochrones suitable for
our purposes (Lejeune \& Schaerer \cite{gen01}, called the Geneva grid
hereafter; Girardi et al. \cite{pad02} or the Padova grid; and Yi et al.
\cite{yi01}, Kim et al. \cite{kim02}, and Yi et al. \cite{yi03}, known as the
$Y^2$ grid). Although the isochrones of the Geneva grid provide the widest
range of ages, they do not show  agreement with present-day observed MSs of
open clusters, whereas the two others do (see Grocholski \& Sarajedini
\cite{grosa03}). The $Y^2$-isochrones  agree well with cluster MSs, and
integrate both Pre- and Post-MS stages.  Unfortunately, they are limited to
small and moderate mass models  ($m \lo 5 m_\odot$), and cannot be applied to
the youngest clusters with  massive and bright stars. Thus we are left with the
Padova Post-MS isochrones, which show good agreement with cluster MSs for $M_V
\lo 7$ mag and are finely spaced over the age scale ($\Delta\log t = 0.05$).
The major inconvenience of the Padova grid is the relatively high lower limit
of the age scale ($\log t =6.6$), which restricts a proper dating of young
clusters. We implement the Padova overshooting isochrone grid with the
following parameters: $m=0.15...66\, m_\odot, Z=0.019, Y=0.273$.

The Pre-MS isochrones were derived by the use of the Grenoble Pre-MS tracks
Internet-server of Siess et al.~(\cite{siess00}). We computed a grid of the
Pre-MS isochrones which covers the same scale of ages as the Post-MS grid.
Their agreement with the Padova models is acceptable and suits our purposes.
So, in this paper we use models with overshooting for a mass range $0.1...7.0\,
m_\odot$ and $Z=0.02, Y=0.28$.

Since the turn-on points are observed only in a few clusters of our sample, we
do not use the Pre-MS isochrones for cluster age determination but show them in
the cluster CMDs to illustrate the degree of conformity of the nuclear and
thermal age scales. Further, the  Pre-MS grid was already  used in this study
for photometric selection of cluster members (see Paper~I for details).

\subsection{The method}

We applied a simple logic for the evaluation of cluster ages:  the age of a
cluster is defined by the average age of individual MS cluster members. In
contrast to the standard method of isochrone fitting, this approach does not
only yield objective estimates of the cluster age but also of its uncertainty
(the standard deviation). Moreover, the algorithm is coded easily and can be
used in the data processing pipeline. Along with the iterations  of membership
determination, the age evaluation procedure can be run in parallel. This is an
important feature because  one is confronted with hundreds of clusters in a
wide range of ages.

The individual ages of stars are derived  from their locations in the CMD with
respect to the isochrone grid (see Fig.~\ref{agev_fig}). Only the most probable
kinematic members are considered. Further, in order to avoid additional
uncertainties which could be introduced by age estimates of red giants (e.g.,
due to the  treatment of convection, mass loss, insecure conversion to the
observed passbands), we restrict the age evaluation to MS stars\footnote{In a
few cases, however, we were forced to use Post-MS stars. Then we selected only
those of them, which are located in the blue part of the  Hertzsprung gap.}.
So, we choose a red border defined  by the Termination Age Main Sequence (TAMS)
which is shown in Fig.~\ref{agev_fig}a as the thick curve most to the right.  
On the other hand, a relatively slow evolution close to  the ZAMS leads to a
high density of the isochrones in this region of the CMD and, consequently, to
a low  accuracy of the interpolation. This forces us to use only an
evolutionary advanced region of the MS where the isochrone crowding effect is
much weaker. This ``blue'' border is shown as the curve to the right of the
ZAMS and is defined below. We call the region of the cluster MS between  the
red and blue borders the Evolved Main Sequence (EMS).

In order to define the blue border of the EMS, we introduce a parameter $s$ as a 
measure of the evolutionary advancement of a given star (called hereafter the
evolutionary status of the star). For a star with absolute magnitude $M_V$ and
colour index $(B-V)_0$, the parameter $s$ is defined as 
\[ 
s=\frac{(B-V)_0-(B-V)_{ZAMS}}{(B-V)_{TAMS}-(B-V)_{ZAMS}}\,, 
\]
where $(B-V)_{ZAMS}$ and $(B-V)_{TAMS}$ are the colour indices of  stars of the
same magnitude $M_V$ on the ZAMS and TAMS, respectively. Since the ZAMS is not
explicitly present in the Padova grid, we adopt the youngest isochrone of $\log
t =6.6$ to represent the ZAMS. At $M_V > -3$ this isochrone coincides well with
the empirical ZAMS of Schmidt-Kaler (\cite{schkal}). 

In Fig.~\ref{agev_fig}b we show age profiles of MS stars for three different 
absolute magnitudes. The profiles have been constructed by spline
interpolations  along the isochrones of different ages. As one can see, at
small $s$ (near the ZAMS) the profiles steepen considerably, making the age
uncertainties due to typical  photometric errors unacceptably large. The
analysis of the MS band over its full length ($M_V =-8...+5$) has shown that
for a relatively safe age determination one should consider stars with $s >
0.15$. For these stars, the ages are determined from the age profiles
corresponding to their absolute magnitudes $M_V$ and evolutionary status $s$.

In order to define a range of absolute magnitudes $\Delta M_V$ where the age
evaluation seems to be reasonable for a given cluster, we  estimate the age of
the brightest member among the evolved MS cluster stars and consider the
corresponding isochrone. Then, the bright limit of $\Delta M_V$ is the absolute
magnitude of this member, and the faint limit is defined as the cross-point
between the corresponding isochrone and the blue edge of the evolved MS. If the
brightest member falls just in the TAMS, then the full spread of the evolved MS
is available for the age calculation. On the other hand, if the brightest
member is located near the blue edge  (i.e., $\Delta M_V \approx 0$), no other
stars are included in the cluster age determination. The distribution of
clusters over the number of the most probable kinematic members used for the
computing of cluster ages is given in Table~\ref{NA_tbl}.

%
\begin{table}
\centering
\caption[]{Distribution of clusters according to the number of members ($N_t$) 
used for age determination. Typically, the more stars could be used the more
reliable is the age determination. The cases $i$ to $iii$ are explained in the
text.}
\label{NA_tbl}
\begin{tabular}{rrrrrr}
\hline
$N_t$&   N clusters & $N_t$ &   N clusters& $N_t$ &   N clusters\\
      &              &        &           &        &              \\
\hline
$> 10$ &   21        &   6    &   17      &   1    &  189         \\
 10    &    5        &   5    &   14      &  $i$   &   47         \\
  9    &    6        &   4    &   29      & $ii$   &    5         \\
  8    &    7        &   3    &   56      &$iii$   &   25         \\
  7    &    3        &   2    &   96      &        &              \\
\hline
\end{tabular}
\end{table}

In Table~\ref{NA_tbl}, there are three special cases marked by $i$, $ii$, or
$iii$ where the applied procedure of age determination does not work. The first
two groups include young clusters whereas the third one consists mainly of old
clusters. For a remote young cluster, the \ascc contains only the top of the
cluster MS  which is generally badly populated due to the IMF depletion in the
domain of  massive stars. For these clusters the Padova grid could not be used 
for our purpose, because for  stars with $M_V \lo -3$ mag, the computed  strip
of the evolved MS becomes  artificially narrow. On the other hand, different
effects like  variable extinction, stellar binarity, rotation etc. increase the
observed  spread of stars around the MS. Therefore, even small photometric
effects could  lead to a situation that no members of remote young clusters
would be present in the narrowed EMS area.  In total, 53 of the young clusters 
do not fulfill the criteria set by the procedure. As a compromise, we consider
the brightest proper motion members approaching the EMS-area either from the 
blue  (group $i$ in Table~\ref{NA_tbl}) or from the red ($ii$) direction and
treated  them in the procedure of age determination as they were just located
at the EMS borders (usually, a shift of less than 0.05~mag parallel to the
$(B-V)_0$ axis  is considered  acceptable). This means that the cluster ages
can be overestimated for the group $i$ or underestimated for the group $ii$.
Wherever it is possible, we prefer the red-edge ages as less affected by random
errors.

For 14 old clusters of the third group ($iii$) we found that their turn-off 
points are fainter than the limiting magnitude of the \ascc. Since  only
sub-giants or red giants in these clusters are present in the \ascc, we are 
not able to determine their age, so we accept, instead, the published values.
For 11 other clusters, the turn-off points are still seen at the faint \ascc
limit but  due to large photometric errors at these magnitudes, the observed
CDMs are very vague. Therefore, ages were estimated manually by isochrone
fitting. Both cases fall into the group ($iii$). The remarks on these special
cases  and decisions made in estimating cluster ages are given in the notes
file added to the \clucat. In order to illustrate the conformance of the
derived ages with the structures of observed CMDs, the corresponding isochrones
are shown in the CMDs of the Cluster Diagram Atlas.

In total, the ages were determined for 506 out of 520 clusters of our sample,
and for 196 of them age estimates are given for the first time.

\subsection{Accuracy of the results and comparison with another scale of cluster
ages}

Although the proposed method of individual age determination is sufficiently
flexible and can be applied to poorly populated and sparse CMDs, there are
several sources of uncertainty which could considerably affect the results. We
can divide them into random errors and biases. Random errors create a
scattering in a cluster CMD  and stem from random photometric errors, variable
extinction, stellar spots etc. Some biases like stellar rotation or unresolved
multiplicity have  strong random parts like the orientation of rotation axes,
or the actual distribution of the component masses. They produce a quasi-random
scattering of stars in the CMDs. The biases due to the adopted cluster distance
and average reddening affect systematically the age estimates of one particular
cluster but they  act more or less randomly if we consider a large sample of
clusters. Finally, there are purely systematic effects which influence the age
estimates of all clusters in the same way, e.g. uncertainties of  the  model
grid which arise basically from the underlying physics or from  neglecting
particular evolutionary phases.

The uncertainty of a random or quasi-random impact on the cluster age
determination can be estimated from the age--absolute magnitude relation (AMR)
at the red edge of the EMS. Since the general shape of the AMR is similar at
the red and blue edges, one could, in principle, consider the blue AMR edge,
too. But due to technical reasons like an insufficient accuracy of published
input data and slow evolution near the ZAMS, the red edge of the AMR suits
better to our purposes. From $rms$ errors in absolute magnitudes
$\sigma_{M_V}$, the age  uncertainty $\sigma_{\log t}$ can be evaluated as 
$\sigma^2_{\log t} = \gamma^2\,\sigma^2_{M_V}$. Here $\gamma$ is the AMR slope
$\gamma = d\log t/dM_V$. For the  Padova isochrones, this slope varies from 0.1
to 0.4 within the complete MS range ($M_V=-8,+4$ mag). In further estimates we
use an averaged slope of $\gamma_{MS}=0.26\pm0.10$.

Let us first estimate the impact of uncertainties in cluster distances onto the
derived cluster ages. The effects we consider include uncertainties in 
distance modulus and reddening. Generally, the clusters of our sample are
located within 1.5~kpc from the Sun, i.e. at galactocentric distances where no
substantial radial metallicity gradient has been detected (see e.g. Andrievsky
et al.~\cite{andr}). Thus, we can adopt the solar abundance of heavy elements
in clusters of our sample and neglect  metallicity corrections. A spread of
cluster metallicity could only arise from a Galactic disk inhomogeneity which
according to Vereshchagin \& Piskunov~(\cite{verpis}), is about
$\Delta\mathrm{[Fe/H]}\approx 0.1$, or $\sigma_{M_V} \approx 0.1$ mag if
converted to absolute magnitudes. Further, a typical accuracy of about 0.2 mag
is expected for a cluster distance modulus (Subramaniam \& Sagar \cite{ss99}),
and a reddening uncertainty  of a few hundredths of a magnitude is derived from
the LGM data. Then, the total $rms$ error due to uncertainties in metallicity, 
interstellar extinction and distance modulus does not exceed 0.3 mag. The
corresponding age error is $\sigma_{\log t} \approx 0.08$.

\begin{figure}[t]
\resizebox{\hsize}{!}
{\includegraphics[bbllx=90,bblly=210,bburx=513,bbury=630,clip=]
                 {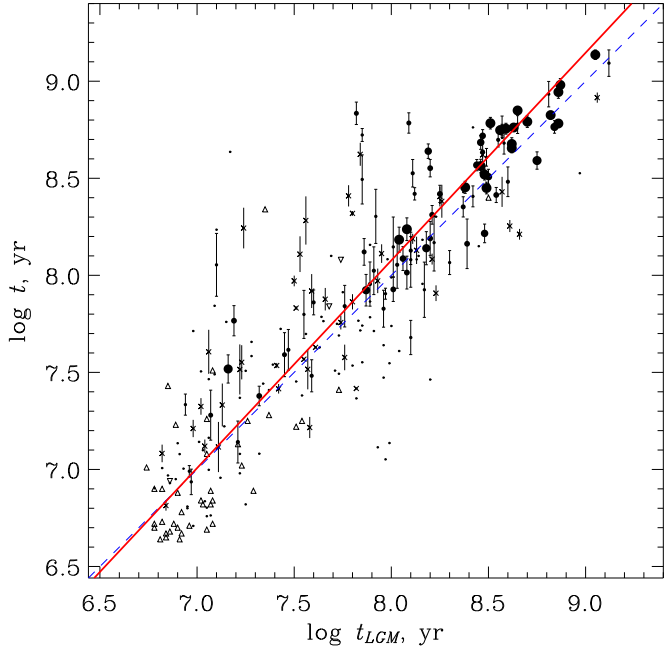}}
\caption{Comparison of the derived cluster ages with data published in LGM.
Dots indicate clusters where only  one star ($N_t=1$) is used for the age
determination, whereas crosses and filled circles are for clusters with $N_t=2$
and $N_t>2$, respectively. The circle sizes are scaled by the number of stars
used for the age determination of a given cluster (cf. Table~\ref{NA_tbl}). For
clusters with $N_t>2$, the bars show $rms$-errors of the averaged ages derived
in this paper. For clusters with $N_t=2$, the bars indicate the scatter and are
shown without hats. Triangles are the clusters of group $(i)$, whereas the
upside-down triangles mark the group $(ii)$. The dashed line is the locus of
equal ages. The solid line shows a regression solution obtained for the
complete sample.} 
\label{agecomp_fig}
\end{figure}

The unresolved multiplicity of MS stars is another effect which may influence
the age determination.  For close unresolved binaries (UBs) an offset of up to
$\Delta M_V= 0.75$ mag exists compared to single stars. At first
glance, the displacement produced by an unresolved component introduces a 
systematic error in the cluster age determination because it always shifts the 
UB up and to the right in the CMD. For the adopted technique of averaging individual
ages it could be regarded as a random effect, however, with no strong bias
component. Indeed, in the vicinity of the MS turn-off/turn-on points, the loci
of single stars and UBs of the same age do cross. This produces a symmetry in
the UBs spread around the isochrone of a single star (see Fig.~\ref{agev_fig}a). In
the plot we compare two isochrones corresponding to populations of single stars and
UBs in a cluster with an age of $\log t= 8.6$. To show the effect at its largest 
amplitude we use the extreme case of components of equal mass.
The resulting symmetry of the UB locus with respect to the
crossing-point (B) in Fig.~\ref{agev_fig}a is evident. One can
see, that unresolved binaries being treated as single stars produce ages both
lower than the cluster age (those residing above point B) and
higher than the true cluster age (below point B). Since we consider in our
approach only the range of evolved stars (those brighter than $M_V(X)$), the
mentioned symmetry is not violated by fainter unevolved UBs. Thus, no
systematic age bias is expected from the unresolved binaries effect in our 
dating technique. On the contrary, for the isochrone-based dating this
effect carries a potential danger: ages are underestimated, when
evolved bright binaries are treated as single stars. Since $\Delta M_V= 0.75$ is
the maximum magnitude spread produced by UBs in the vicinity of the single-star
turn-off point, we estimate that this effect introduces $\sigma_{\log t} < 0.19$
to the average cluster age.

We do not consider the impact of mass loss onto the age determination since
this  effect is implicitly taken into account by the isochrones involved.
Although stellar rotation can change the observed absolute magnitude of the MS
stars, the effect is less important than the one introduced by unresolved
multiples. According to Maeder \& Peyntremann~(\cite{maed70}), for critically
rotating stars (near the break-up velocity) with masses of $1.4-5\, m_\odot$,
the offset in brightness is less than $\pm0.6$~mag (i.e. $\sigma_{\log t} \le
0.16$). This effect decreases rapidly with slowing down the rotation, and at a
rotation velocity of 0.8 of the critical one the offset is $|\Delta M_V| \lo
0.2$ ($\sigma_{\log t} \le 0.05$). Due to the random orientation of the aspect
angle it has a random character.

Also, the uncertainties of the age calibration due to the hooks of the
isochrones near the red border of the EMS are small, since the time of the
overall contraction after the hydrogen core exhaustion is relatively short.
According to the simulations which take into account this evolutionary phase,
$\log t$ of an individual star within the hook area  changes by less than 0.02.

We conclude that the accuracy of age determination for clusters in our sample
is about $\sigma_{\log t} = 0.20 - 0.25$ and compare the results  with uniform
data on cluster ages recently published by LGM. The cluster ages in LGM are
derived via the method of isochrone fitting and they have the same theoretical
basis (i.e. the Padova grid). The authors make use of photoelectric $UBV$
observations without an explicit selection of cluster members. Totally, the LGM
and our samples have 255 clusters in common, with 52 of them in the groups
$(i)$ and $(ii)$ (cf. Table~\ref{NA_tbl}). Since both methods use the same isochrone grid,
the ages of a single cluster differ only due to the applied algorithms and the
selected  observations.

The results of the comparison are presented in Fig.~\ref{agecomp_fig}.  As both
coordinates are subject to $rms$ errors, a simple linear regression analysis is
inappropriate. A least square bisector solution as proposed by Isobe et
al.~(\cite{iso90}) should rather be adopted. For all clusters in common, this
solution of the equation
\[
\log t = c_0 + c_1\cdot\log t_{LGM},  
\]
yields $(c_0,c_1)$=$(-0.48\pm0.18,1.07\pm 0.02)$. It is shown in
Fig.~\ref{agecomp_fig} as a solid line.

The mean age difference turns out to be  $\overline{\Delta\log
t}$=$-0.06\pm0.02$, which is small compared to the standard deviation of
$\sigma_{\Delta\log t}$=$0.32$. According to  Fig.~\ref{agecomp_fig}, a
relatively large spread is caused by 27 outstanding clusters above the
regression line and 4 clusters below this line. Checking these clusters in
detail, we found that only for a few very distant clusters the large deviations
can be explained by the low accuracy of the cluster CMDs near the limiting
magnitude of the \ascc.  For the majority, the different stellar content (i.e.
adopted members) between LGM and this paper is the decisive factor.  In case of
clusters above the regression line, bright stars in the cluster  areas were
considered by LGM as cluster members, but according to their kinematics we
found that they are non-members. Therefore, we may assume an underestimation of
the true ages for these clusters by LGM rather than an overestimation by us. In
the case of the clusters below the regression line, the effect is reversed:
bright stars  within the central cluster area were rejected by LGM as
non-members,  although they fulfill all selection criteria we adopt for
members.

Excluding these 31 clusters from consideration, we obtain for the mean age
difference and its standard deviation $\overline{\Delta\log t}$=$0.01\pm0.01$
and $\sigma_{\Delta\log t}$=$0.20$, respectively. We may conclude that the
applied technique allows to derive age estimates with an accuracy which is
comparable with the accuracy of the classical method of isochrone fitting.
According to the discussion above, the small but significant systematic 
bias ($c_1=1.07$) can be explained by a slight underestimation of the LGM 
isochronic ages of older ($\log t > 8.3$) clusters due to the UB effect.
On the other hand, since the lower limit of the Padova isochrones is set
at $\log t=6.6$, this yields a bias by somewhat ``compressing'' the left
side of the above relation and might increase $c_1$, too.

\section{Conclusions}\label{concl_sec}

Starting from a homogeneous sky-survey, we determined a number of astrophysical
parameters of Galactic open clusters. This sky survey, i.e. the \ascc, allowed
us to perform an unbiased rediscussion of the membership in clusters
(Kharchenko et al.~\cite{starcat}), a necessary requirement for the
characterisation of the clusters. The results of this paper are published in
the form of an open cluster catalogue, which contains for all of the 520
clusters investigated: heliocentric distance, interstellar extinction along the
line of sight, angular size, mean proper motion and radial velocity, and age.

For 200 clusters heliocentric distances have been newly determined via
MS-fitting. We calibrated our results by a comparison with a subset of clusters
in Loktin's (Loktin et al.~\cite{lok01}, \cite{lok04}) sample, distributed over
a wide range of distances.

For 301 clusters mean proper motions in the Hipparcos system have been derived
for the first time. They are based on the individual proper motions of the most
probable members. For the remaining clusters proper motions are compared with
the results of Robichon et al.~(\cite{rob99}), Baumgardt et
al.~(\cite{baum00}), and Dias et al.~(\cite{dla1}, \cite{dla2}) and good
agreement within a few mas/year has been found.

For 290 clusters in our sample, mean radial velocities could be determined
based on identifications of our members in the catalogue of Barbier-Brossat \&
Figon~(\cite{gcrv}), or retrieved from the literature. For 94 clusters radial
velocities are determined for the first time, for others our findings were
compared with Lyng{\aa}~(\cite{lyn87}), Ruprecht et al. (\cite{rupr}), and
DLAM, and good agreement was found. On the other hand, there are 230 clusters
in our sample, for which not even a single radial velocity measurement of any
member is available so far.

Considerable effort has been put into the determination of cluster ages. In the
end, ages could be determined for 506 out of 520 clusters, of which 196 are
first estimates. We compared our determinations with the work of Loktin et
al.(~\cite{lok01}, \cite{lok04}) and found good agreement. For 31 cases with
large discrepancies between this paper and LGM, these can be explained by the
different membership criteria between the two papers. 

Angular sizes of the cluster cores and coronas have been newly determined. This
is the area where the full-sky coverage of the \ascc is very helpful. It makes 
the size determination free from selection effects, such as limited field, star
counts or visual inspection of the sky without taking into account membership
criteria, and others.  In this paper we proceeded the following way: from our
new membership list, the new centre and the radial density distribution
function were determined. Then by visual inspection of each of the clusters,
core and coronas radii were determined. Despite of the relatively bright
limiting magnitude of the sky survey we used (\ascc with a completeness limit
at $V = 11.5$ and a limiting magnitude at $V = 14.0$), we find that the angular
sizes of the cluster coronas are systematically larger than in earlier
determinations in the literature. If mass segregation in open clusters is an
important issue (see e.g. de~Grijs et al. \cite{griea02}), our determinations
of the cluster coronas are probably only lower limits of the true sizes.

To improve this situation, a homogeneous sky survey with fainter limiting
magnitude is needed. Although such surveys are available in infrared photometry
(e.g. 2MASS), a survey is missing in the optical and ultraviolet regimes, it is
missing as far as proper motions are concerned, and it certainly cannot be
expected soon for radial velocities. More than a decade from now, the ESA
project GAIA is supposed to provide a survey which will fulfill all the
requirements stated above. In the meantime, progress has to be made in the
field of ground-based astronomy in order to derive a homogeneous, bias-free 
survey in multicolour photometry (SEGUE in the Galactic plane), in proper
motions (beyond UCAC), and in radial velocities (RAVE).

\begin{acknowledgements} 
This work was supported by the DFG grant 436~RUS~113/757/0-1, RFBR grant
03-02-04028, and by the FCNTP "Astronomy". We acknowledge the use of the Simbad
database  and the VizieR Catalogue Service operated at the  CDS, France, and
WEBDA facility at the Observatory of Geneva, Swiss. We thank the referee, J.-C.
Mermilliod, whose comments helped us to improve the paper.
\end{acknowledgements}

\appendix

\section{The Catalogue of Open Cluster Data}\label{cat_sec}

The Catalogue of Open Cluster Data (\clucat)  exists in machine readable form
only and can be retrieved from the CDS online
archive.\footnote{\texttt{ftp:cdsarc.u-strasbg.fr/pub/cats,
http://vizier.u\mbox{-}strasbg.fr}} The catalogue consists of three files:
format description (ReadMe), the main table with the derived parameters and
literature data, and a notes file. The necessary details on the data format can
be found in the ReadMe file. In order to inform the reader on the data scope
included in the catalogue, we describe here the main table. The table contains
data on 520 clusters and consists of 520 lines, and 30 columns. The column
description is given in Table~\ref{catcon_tab}.

\begin{table}[t]
\caption{Contents of the \clucat main table}
\label{catcon_tab}
\setlength{\tabcolsep}{2pt}
\begin{tabular}{rlcl}
\hline
Col& Label&Units&Explanations\\
\hline
   &&&\\
1  &No            &  ---   & Sequential number\\
2  &Name  &  ---   & NGC, IC or other common\\
   &              &        & designation\\
3  &RA            &   h    & RA J2000.0 of the center\\
4  &Dec           &  deg   & Dec J2000.0 of the center\\
5  &l             &  deg   & Galactic longitude of the center\\
6  &b             &  deg   & Galactic latitude of the center\\
7  &Rco           &  deg   & Angular radius of the core\\
8  &Rcl           &  deg   & Angular radius of the cluster\\
9  &RV            &  km/s  & Average radial velocity\\
10 &eRV           &  km/s  & Error in RV\\
11 &nRV           &  ---   & Number of stars used for RV\\
   &              &        & calculation\\
12 &PMx           & mas/yr & $\mu_\alpha\cos\delta$: average proper\\
   &              &        & motion in RA \\
13 &ePMx          & mas/yr & Error in PMx\\
14 &PMy           & mas/yr & $\mu_\delta$: average proper\\
   &              &        & motion in Dec\\
15 &ePMy          & mas/yr & Error in PMy\\
16 &PMl           & mas/yr & $\mu_l\cos b$: average proper\\
   &              &        & motion in l\\
17 &PMb           & mas/yr & $\mu_b$: average proper motion in b\\
18 &N1s           & ---    & Number of most probable ($1\sigma$)\\
   &              &        & members\\
19 &d             & pc     & Distance from the Sun\\
20 &E(B-V)        & mag    & Reddening\\
21 &V-Mv          & mag    & Apparent distance modulus\\
22 &source(d)     & ---    & Source of distance and $E(B-V)$\\
23 &logt          & log yr & Logarithm of average age\\
24 &Nt            & ---    & Number of stars used \\
   &              &        & for the calculation of logt\\
25 &RVref         & km/s   & RV from literature\\
26 &eRVref        & km/s   & Error of RVref\\
27 &source(RVref) & ---    & Source of RVref\\
28 &logtref       & log yr & $\log t$ from literature\\
29 &source(logtref)& ---   & Source of logtref\\
30 &note flag     & ---    & References to note file\\
&&&\\
\hline
\end{tabular}\\
\end{table}

\section{The Open Cluster Diagrams Atlas}\label{atl_sec}

The Atlas (OCDA) presents visual information on the data used for the 
determination of cluster parameters, on the quality of member selection, and 
on the accuracy of the derived parameters. The OCDA consists of 520 PostScript 
plots stored as gzipped files (i.e. one file per cluster), which will   be
available in electronic form only via the CDS.  In order to get an easy access
to the catalogue data, the file name includes the  sequential number of the
corresponding cluster in \clucat and the cluster name  (column 2 in the main
table). Each plot contains a header and five diagrams  used in the reduction
pipeline. An example is given in Fig.~\ref{atlas_fig} for the Coma Berenicis
cluster.

\begin{figure*}[t] \resizebox{\hsize}{!}
{\includegraphics[bbllx=30,bblly=30,bburx=565,bbury=795,angle=270,clip=]
{figs/308_Melotte_111.ps.gz}}
\caption{An example of the atlas: A plot for the open cluster Coma Berenicis.
}\label{atlas_fig}
\end{figure*}

In the header we provide the equatorial coordinates of the newly determined
cluster center, the cluster number in the \clucat and the most common
designation of the cluster. The panels present spatial, kinematic and
evolutionary information. The upper row: the left panel is a sky map of the
cluster neighbourhood constructed with stars from the \ascc, and the right 
panel is  the colour-magnitude diagram. The bottom row: the left panel shows
radial profiles of the projected density, the middle  panel is the vector point
diagram of proper motions, and the two right panels are ``magnitude equation'' 
($\mu_{x,y}-V$ relation) diagrams. 

The sky map: A blue cross marks the adopted cluster centre. Pluses are centres
of all clusters located within the displayed area and taken from the
literature, i.e. the large plus in blue is for the given cluster, the smaller
magenta ones (absent in the example) are for other clusters. The large circles
are the borders of the cluster core (solid curve) and of the corona (dashed
curve). The small circles are stars; their size indicates (only in this panel)
stellar magnitude. The bold circles are $1\sigma$ cluster members  ($P_c \ge
61$\%), i.e. black circles are the members in the core area, red circles --
members in the corona. The stars located outside the corona are displayed in
cyan.

The colour-magnitude diagram: In this and the following diagrams stars are
marked as coloured dots. $1\sigma$-members are shown with error bars. Stars
used for the age determination are marked as bold magenta circles. Colours of
the curves: magenta for the adopted Post-MS and Pre-MS isochrones, cyan for the
borders of the Evolved Main Sequence and the ``mean'' location of the
Hertzsprung gap, blue for the borders used in the photometric selection (see
Paper~I for details). Additionally, cluster parameters selected from the
\clucat are plotted where the $rms$ errors are given in parentheses, and the
references to the published data in the brackets. The symbol (\#) in the line
with the average age (log~t) indicates the number of stars used for the age
determination.

The density profiles diagram: The curves show the distributions of stars with angular
distance from the cluster center. The green curve shows all stars in the
cluster area, the magenta and black curves are for $3\sigma$- and $1\sigma$-
members, respectively. The adopted core and corona radii are shown by solid and dashed
lines. 

The vector point diagram shows the proper motion distribution of \ascc stars 
in this sky area. The error bars are shown for the cluster members.

The magnitude equation diagrams show the distribution of the proper motions
versus magnitudes. The horizontal lines correspond to the average proper motion
of the cluster, the bars indicate the $rms$ errors of proper motions for the
cluster members.

\end{document}